\documentclass[fleqn,twoside]{article}
\usepackage[headings]{espcrc2}

\readRCS
$Id: robot.tex,v 1.2 2004/02/24 11:22:11 spepping Exp $
\usepackage{graphicx}
\usepackage[figuresright]{rotating}

\title{A novel automatic film changer for high-speed analysis of nuclear emulsions.}

\author{
         K. Borer\address[LHEP]{Laboratory for High Energy Physics, University of Bern,  Sidlerstrasse 5, Bern, 3012, Switzerland},
	 J. Damet\addressmark[LHEP]\address[LAPP]{Now at LAPP, IN2P3-CNRS and Universit\`{e} de Savoie, Annecy, France},
         M. Hess\addressmark[LHEP],
         I. Kreslo\addressmark[LHEP], 
         U. Moser\addressmark[LHEP],
         K. Pretzl\addressmark[LHEP],
	 N. Savvinov\addressmark[LHEP], 
         H.-U. Sch\"{u}tz\addressmark[LHEP],
	 T. W\"{a}lchli\addressmark[LHEP],
	 M. Weber\addressmark[LHEP]\address[FNAL]{Now at Fermi National Laboratory, Chicago, USA}.
         }

\begin{document}

\begin{abstract}
This paper describes the recent development of a novel automatic computer-controlled manipulator for emulsion film placement and removal at the microscope object table (also called stage). 
The manipulator is designed for mass scanning of emulsion films for the OPERA neutrino 
oscillation experiment and provides emulsion changing time shorter than 30 seconds with an emulsion film positioning 
accuracy as good as $20$~$\mu$m RMS.

\vspace{1pc}
\end{abstract}

\maketitle

\section{Introduction}

The extensive use of nuclear emulsions as precise tracking detectors in experimental physics has been made possible  due to recent 
advances in the production of novel nuclear emulsion films and to the development of automatic scanning devices. 
The emulsions with dimensions as small as $12.5\times10$~cm$^2$, as used in the CERN-LNGS OPERA experiment,
are commercially produced by the Fuji Film\footnote{Fuji Film, Minamiashigara, 250-0193, Japan} 
company.
The scanning speed of such scanning devices has reached 20~cm$^2$ of emulsion surface per hour. 
However, so far the emulsion films were fed to the microscope by the hands of an operator. Given the present scanning speed the development of an automatic emulsion film changing system has become mandatory.

\section{Motivation and requirements}
\subsection{The OPERA neutrino oscillation experiment}
The direct observation of $\nu_{\mu} - \nu_{\tau}$ oscillations in 
a $\nu_{\mu}$ beam 
is the main goal of the OPERA experiment \cite{Proposal}. The neutrino beam is provided by the CERN CNGS facility.
The detector is placed at the distance of 732~km in the underground hall of the Laboratory of Gran-Sasso (LNGS).
The Emulsion Cloud Chambers (further - ECC) technique is used to precisely reconstruct the topology of $\tau$ decays produced in
$\nu_{\tau}$ CC interactions with a massive lead/nuclear emulsions target.

\begin{figure}
\includegraphics[width=.5\textwidth]{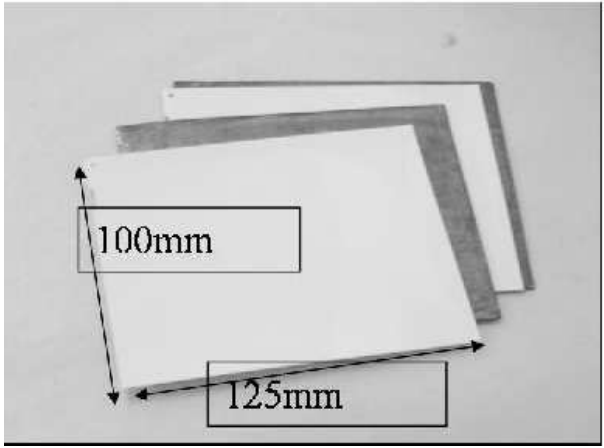}
\caption{Emulsion films and lead plates used for the ECC bricks of the OPERA experiment.}
\label{fig:Emu}
\end{figure}

The OPERA detector has a mass of 1.8~kton and consists of a lead/emulsion film sandwiched target, a target tracker to
localize the event within the target and a muon spectrometer\cite{Proposal}.
The OPERA target is composed of about 200000 ECC bricks of $12.7~\times~10.2~\times~7.5$~cm$^2$ each ,
resulting in 11.4~million emulsion films (see Figure~\ref{fig:Emu}) with a total surface area exceeding 130000~m$^2$.

The expected number of bricks to be processed is about 20 per day for an average beam intensity 
with peak load of about 50 per day.
Assuming about 20 scanning stations available in the OPERA collaboration, the expected average rate will be 
1 brick per scanning station per day. Scanning of one brick (57 emulsion films) must be done in a few passes. Therefore, the number of times that the emulsion film must be placed onto the table of the microscope may easily exceed 100 per day.
In order to handle this large number of operations the film placement must be automatized.

\subsection{Automatic scanning microscopes}

In the old days emulsion scanning was performed by human operators by means of optical microscopes.
This technique (so called "Eye scan") requires large manpower, operators must be highly trained and qualified
in order to achieve good track recognition efficiency. As the scale of experiments has grown, the necessity of
a computer controlled automatic procedure became evident.

An important step in this direction was made at Nagoya University. The system called Track Selector 
was developed in 1982 and used in WA75, CHORUS\cite{Eskut} and DONUT\cite{Kodama} experiments. It evolved its
scanning ability from  0.2 microscope views per second in 1982  to 30 views per second in 2001 \cite{Aoki,Nakano}.
A similar progress has been made recently in Europe. A system able to scan 20~cm$^2$/hour of emulsion surface with
 real-time track reconstruction has been developed by European groups of the OPERA collaboration \cite{Bozza,Sirri}.
The automatic emulsion film manipulator described in this paper has been designed and built as an add-on to the
existing design of the OPERA European Scanning System (ESS).

\subsection{Requirements: Positioning accuracy}
The reference coordinate system for finding particle tracks in emulsions is based on so-called "fiducial marks" printed on the emulsion surface by a dedicated projecting system.
Such marks can be recognized by the scanning microscope \cite{Sirri} and their position with respect to some arbitrary reference point may be measured with submicron precision.
By measuring a minimum of 3 marks, an unambiguous affine transformation is established between the two coordinate systems.

The mark search is performed along a spiral path with 200~$\mu$m pitch (the microscope view is about $390\times310$~$\mu$m$^2$).
One step on this path takes about 300 ms. Hence, the time needed to find a mark rapidly increases with the distance from the search start point to the mark
position: $T=T_0\cdot(D/P)^2/2$  where $P$=200~$\mu$m is a search step pitch, $T_0$=300 ms is the time needed for one step and $D$ is the distance to the mark. 
The first task of the manipulator is to place the emulsion in such a way that marks will be within one microscope view from their
nominal position, so that they 
will be found within 300 ms each. This implies the requirement of a placement accuracy better than 300~$\mu$m peak-to-peak. 

\subsection{Requirements: Operation speed}

The number of emulsion replacement operations may exceed 100 per day. The time spent for scanning is presently estimated to be about 16 hours
per brick on average.  The minimum time needed to establish the reference coordinate system (scan fiducial marks) is 30 s per emulsion.
Hence, the time that can be spent for one placement operation is limited to 3 minutes per emulsion film for a single replacement.

\subsection{Requirements: Failure rate}
The only repeating failure that is tolerable during the operation of the emulsion manipulator is the emulsion
loss during taking from the microscope table or from the bank.
The system is supposed to perform scanning in automatic mode 
24 hours a day. An emulsion loss would require human intervention that can be reasonably made only during 8 hours a day.
For this reason the rate of such failures must be limited to less than one per 10$^3$ placements in order to ensure the dead time being within $5\%$.

\subsection{Mechanical requirements}
The manipulator must be mounted onto the support of the automated emulsion scanning microscope \cite{Sirri}.
This sets a limit to the tolerable level of vibrations generated by the manipulator. The maximum acceptable horizontal acceleration of the table during emulsion scanning is very low. Since the plate changing is only performed when scanning is finished,
the limit is derived just from the mechanical stability of the system.
Rough estimations result in the requirement that the moving part must not apply an inertial force to the microscope support in excess of 10 kg.

\section{Mechanical design}
\subsection{Support}
\begin{figure*}
\includegraphics[width=.9\textwidth]{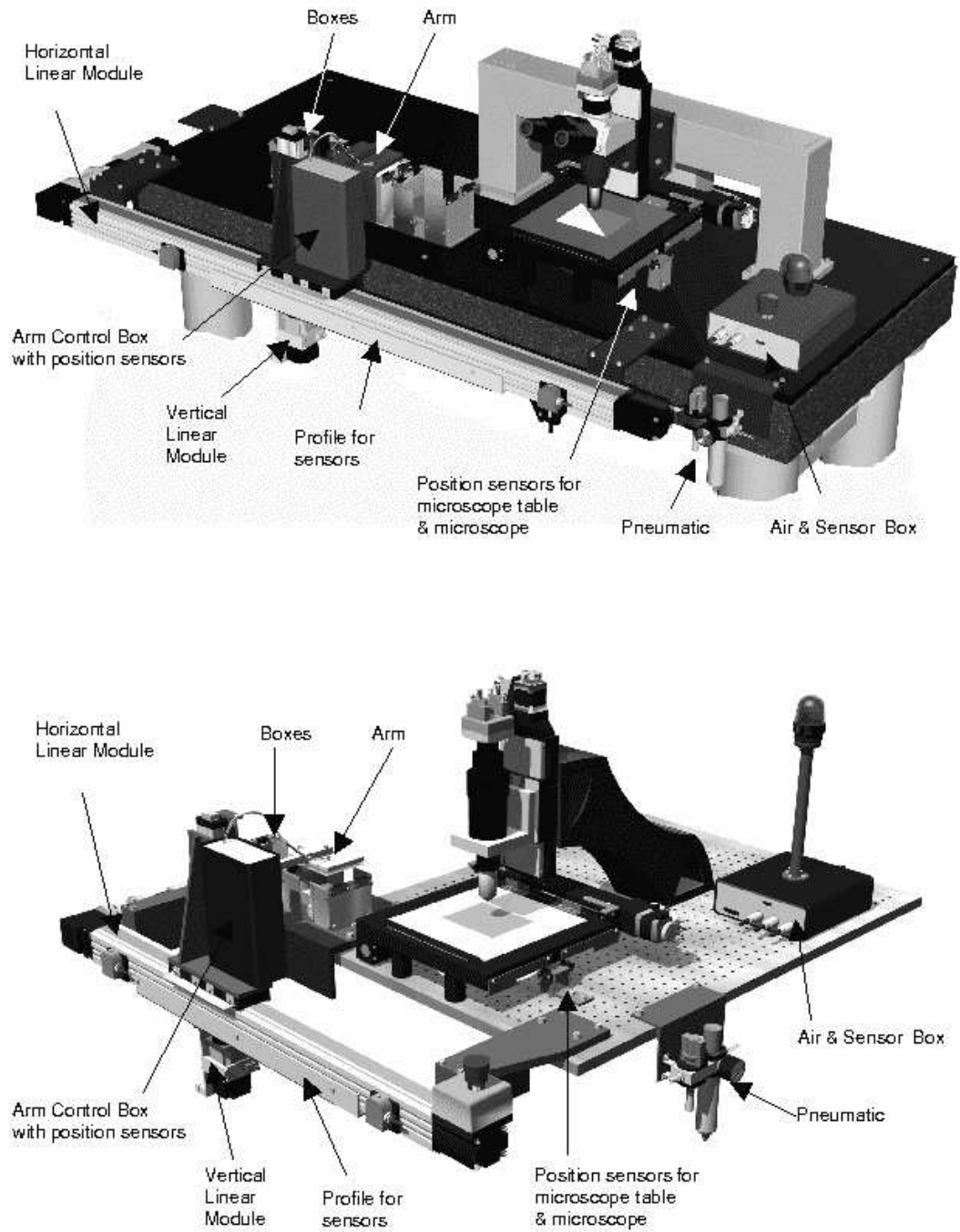}
\caption{Scanning stations equipped with the novel automatic emulsion film manipulator. 
Top - long table design, bottom - short table design.
}
\label{fig:MIL1}
\end{figure*}

The general view of the manipulator mounted on the emulsion scanning microscope table is shown in Figure~\ref{fig:MIL1}.
The manipulator horizontal drive is attached by two aluminum brackets to the microscope ground plate at
the two extreme corners. The OPERA scanning microscopes in Europe have three different designs for this plate.
The shortest one is used by most of the laboratories and is 90 cm wide. This is only enough to accommodate two emulsion boxes, which represents 
the minimal configuration. In some laboratories a longer table design is used (Figure~\ref{fig:MIL1}).  Such table can host up to four bank boxes on it, making loading and unloading of emulsion piles easier. Three emulsion bricks can be scanned simultaneously
without human intervention. 

\subsection{Linear Drives}

The movement of the vacuum arm is performed within two degrees of freedom. Two Rexroth\footnote{Bosch Rexroth Schweiz AG  (Bosch Group),  Hemrietstr. 2,  8863 Buttikon, Switzerland. } linear drives
are utilized for this purpose. The horizontal drive MKR15-65 is fixed directly to the microscope ground plate with two 
brackets. The vertical drive  PSK-50 is mounted onto the carriage of the horizontal drive and moves with it.
The arm is mounted on the carriage of the vertical drive. The horizontal drive positions the vertical drive in front of a given bank box or of a microscope table,
while the vertical drive moves the arm up and down in order to take emulsion films or place them.

%\begin{figure}
%\includegraphics[width=0.5\textwidth]{MIL2-00.01.eps}
%\caption{Manipulator design}
%\label{fig:design1}
%\end{figure}

%\begin{figure}
%\includegraphics[width=0.5\textwidth]{MIL2-00.02.eps}
%\caption{Manipulator design}
%\label{fig:design2}
%\end{figure}

\begin{figure}
\includegraphics[width=0.5\textwidth]{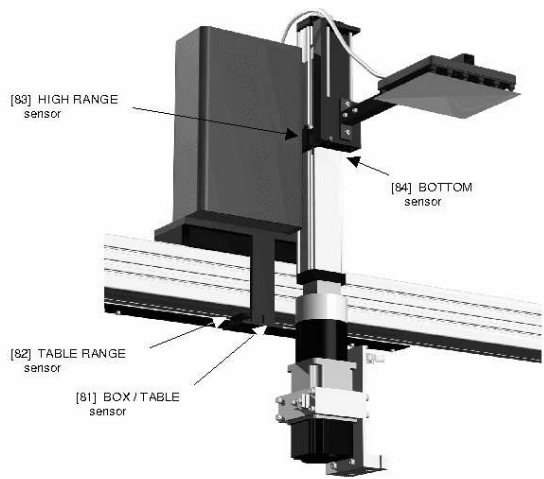}
\caption{Manipulator design: the vertical drive.}
\label{fig:vdrive}
\end{figure}

\subsection{Emulsion bank boxes}
\begin{figure}
\includegraphics[width=0.5\textwidth]{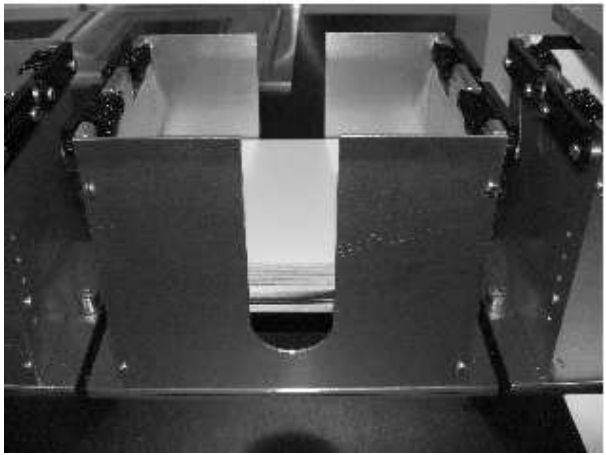}
\caption{Bank box loaded with emulsion films and plastic separators.}
\label{fig:box}
\end{figure}

The emulsion films are supplied to the manipulator from bank boxes located next to the microscope object table.
The number of boxes varies from two to four. Each box is equipped with brushes at the upper edge (see Figure~\ref{fig:box}).
These brushes help the manipulator in separating the target film from the underlying films, which may stick to it.
The robot makes a predefined number of passes across these brushes to ensure reliable separation.

In order to achieve high scanning efficiency the OPERA emulsions have to keep relatively high water content within the gelatin layer, which makes them somewhat sticky.
Therefore it is not possible to pile them up without special low adhesion separating plates. In order to avoid
sticking 0.3 mm thick polystyrene sheets are used as such plates.

\subsection{Vacuum table}

\begin{figure}
\includegraphics[width=0.5\textwidth]{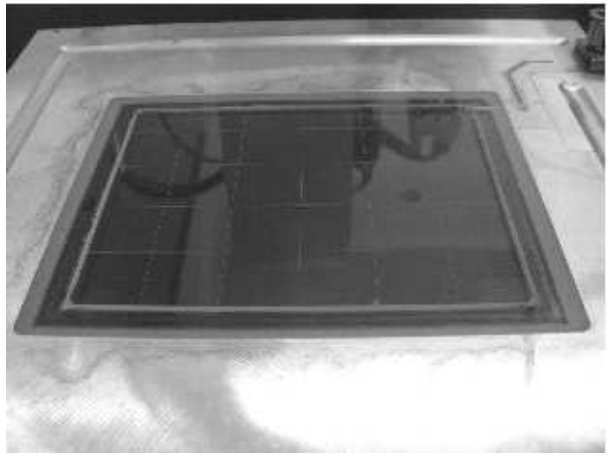}
\caption{Microscope object table.}
\label{fig:table}
\end{figure}

\begin{figure}
\includegraphics[width=0.5\textwidth]{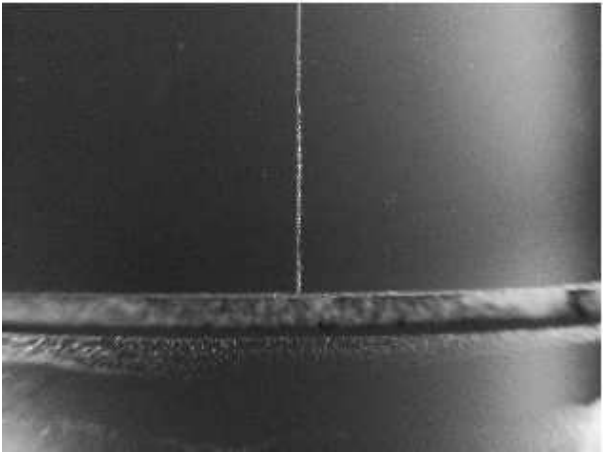}
\caption{Etched groove (vertical) on the glass table.}
\label{fig:groove}
\end{figure}

The object table (see Figure~\ref{fig:table}) has been developed by the INFN Bologna group of OPERA \cite{Sirri}.
In order to avoid  air bubbles between the emulsion and the glass surface a few tiny grooves are
etched on it. They go across the glass part of the table from one side to the other connecting 
the broad mechanically-made vacuum channels (0.5~mm wide) on sides of the plate (see Figure~\ref{fig:groove}). 
The width of these etched grooves is about 50~$\mu$m and the depth 30~$\mu$m.
The etching mask made of an elastic acrylic paint was deposited onto the glass surface and scratched by
a sharp blade along the grooves position.
The grooves are etched in a $20\%$ aqueous solution of ammonium bifluoride at 20$^\circ$C for about 10 minutes.
They allow the air to be removed from below the emulsion within 30 seconds. The scan of
the fiducial marks is not affected by the presence of this air and can be done in parallel.

When emulsion film needs to be taken from the table, a positive air pressure is applied to the groove channel to simplify 
the detachment of the emulsion.
Both vacuum and positive pressure are generated by miniature vacuum pump with ejector pulse VADM-I-70-P by Festo.\footnote{Festo AG
Moosmattstr. 24, 8953 Dietikon / ZH, Switzerland.}
This pump needs a supply of an air with the pressure in the range of 2-8~atm and generates vacuum down to 0.2~atm. The air consumption is less than 20~l/min.

\begin{figure}
\includegraphics[width=0.5\textwidth]{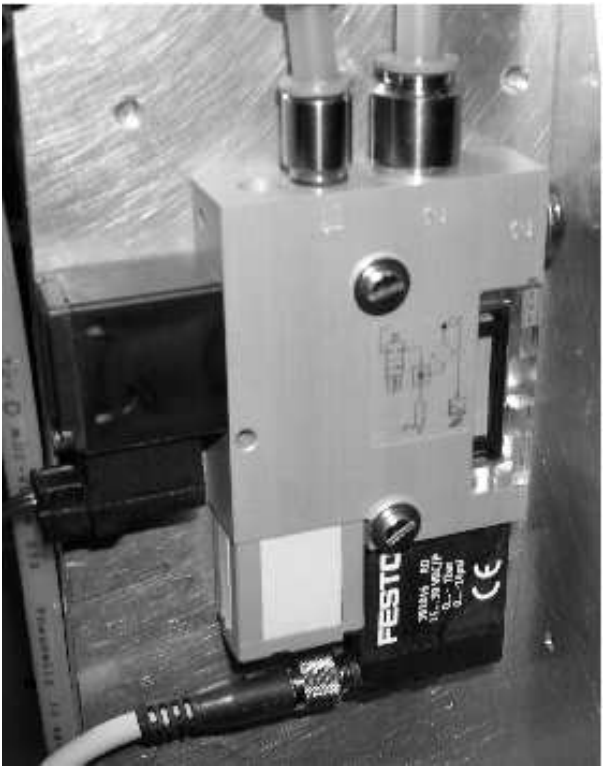}
\caption{Venturi vacuum pump FESTO VADM-70-P.}
\label{fig:festo}
\end{figure}

\subsection{Vacuum arm}
\begin{figure}
\includegraphics[width=0.5\textwidth]{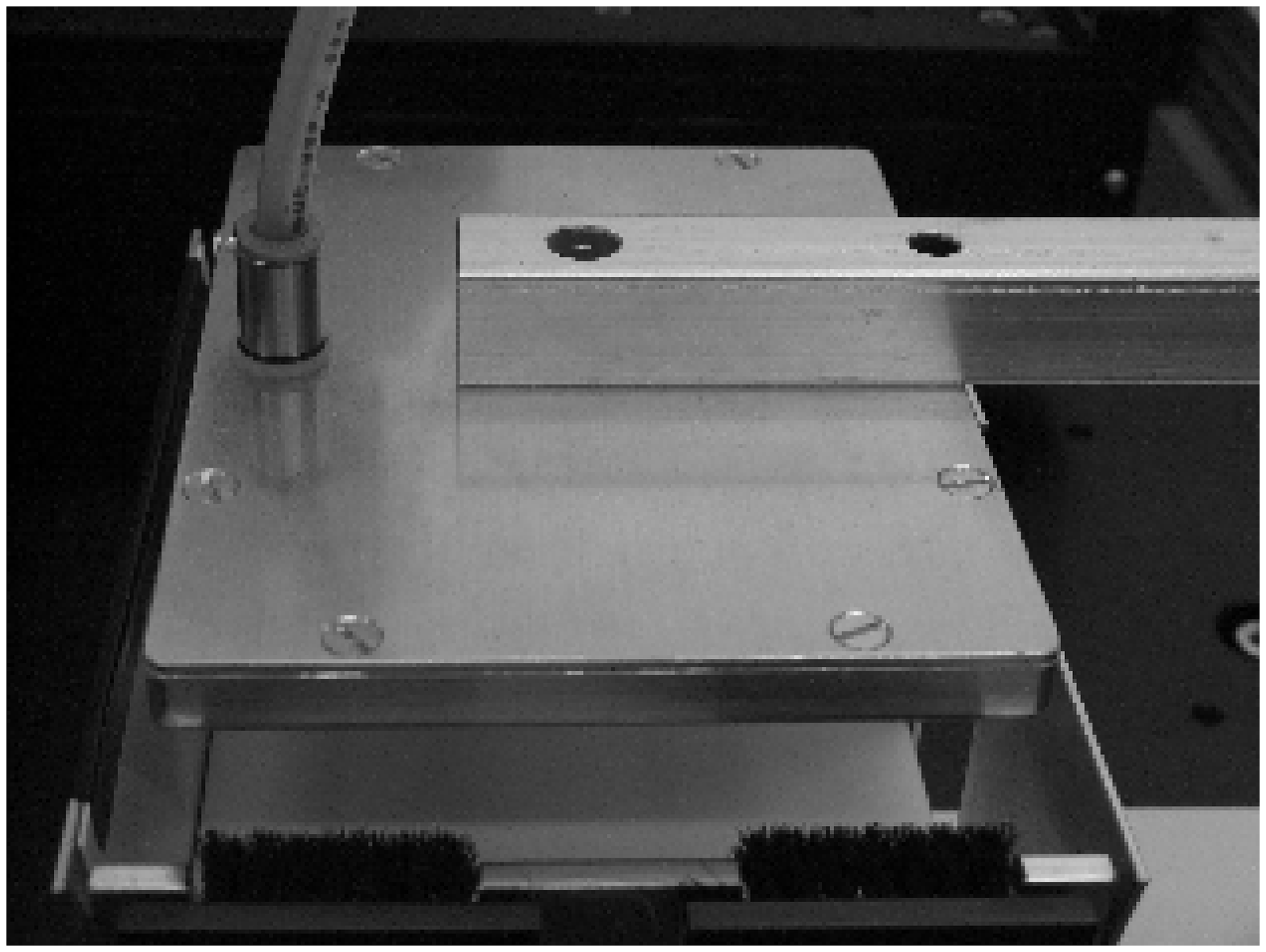}
\caption{Manipulator arm.}
\label{fig:arm}
\end{figure}

The manipulator vacuum arm consists of a yoke and a head. The head hosts 10 vacuum cups (Figure~\ref{fig:arm}) that hold
the emulsion film during arm motion.
The cups are located on the lower surface of the head in such a way that they pick emulsions against the vacuum grove on the microscope object table. This configuration provides reliable emulsion removal from the table.
Vacuum for the arm is generated by another vacuum pump without ejector pulse VADM-70-P by Festo (Figure~\ref{fig:festo}).

\section{Electronics}

\begin{figure*}
\includegraphics[width=0.9\textwidth]{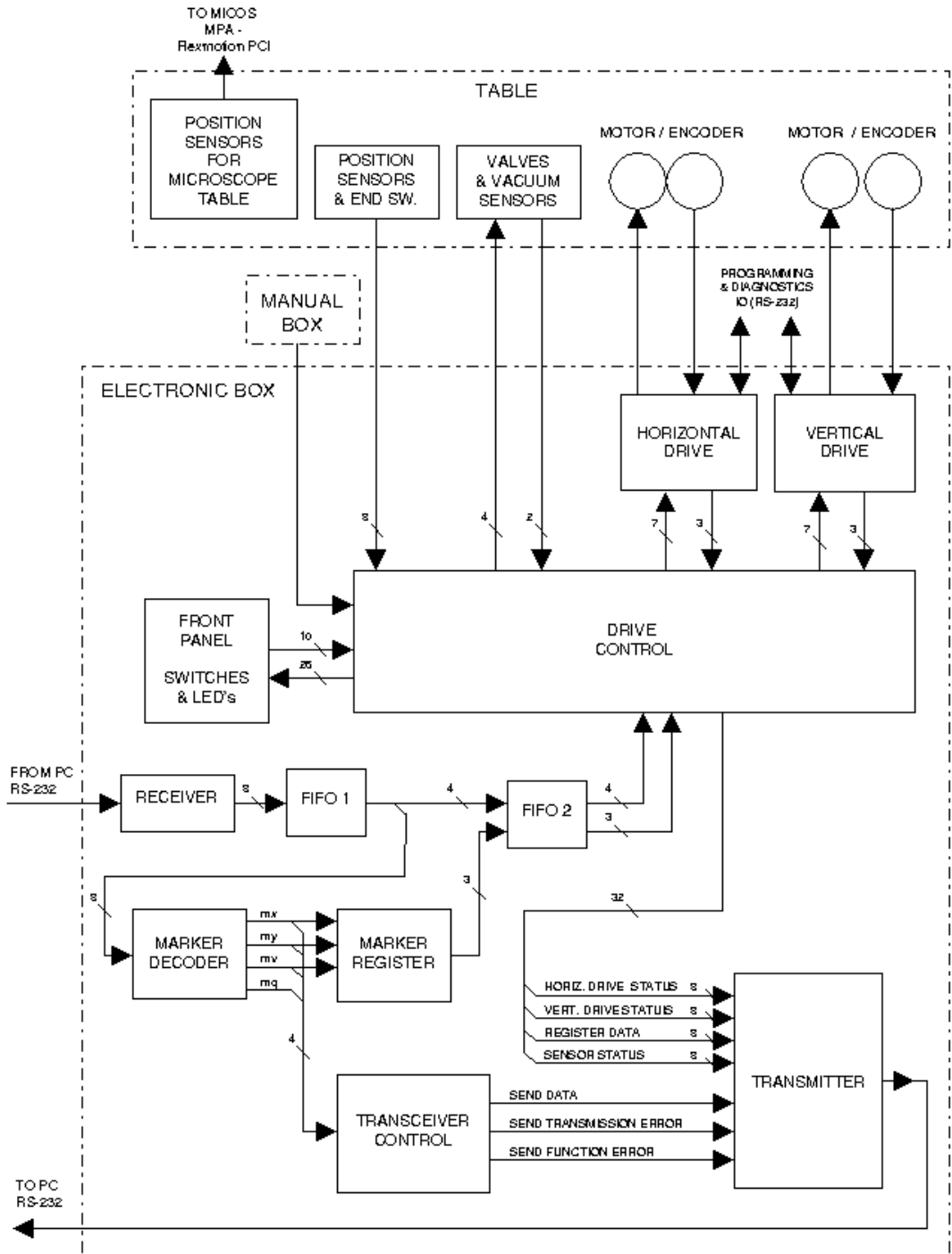}
\caption{Block scheme of electronic control module of the manipulator.}
\label{fig:electro}
\end{figure*}
The block scheme of the manipulator control electronics is shown in Figure~\ref{fig:electro}.
The main control module is realized in a standard 19" 6U rack mounted case.
It is connected to the microscope table and to the manual control box by a 
series of cables.
The RS232 transceiver and the main control logics are realized in Altera FLEX10K FPGA.
The transceiver has a 2-level FIFO for 2$\times$16 commands, which are executed sequentially.
The only exception is the "Q" command: "Query system status". This command is not queued
but executed immediately allowing to make real-time status polling.

The power motors of linear actuators are driven by Rexroth Drive Control modules.
In order to avoid any possible mechanical damage of the manipulator structure due to improper movements
(for example horizontal motion when the arm is in low position may break the arm) a set
of sensors is installed on both horizontal and vertical  drives. Signals from these sensors are used
by interlock logics to prevent such a risk. The vertical motion of the arm
is only permitted when the horizontal position matches film boxes or the microscope table.
In turn the horizontal motion is only allowed when the arm is in the top position. 

The manual control module has four buttons to slowly move the arm in both horizontal and vertical directions.
To move fast to the predefined position the position number must be set by the dedicated switches.
Pressing the "GO" button  then executes the motion.
The arm and table vacuum and pressure switches are mounted at this module as well.
The manual control is activated by a switch, which at the same time disables control signals from the RS232 line.

In order to improve the operation safety, sound and light alarms located at the microscope table are activated
when the arm is about to move. The delay between the warning and the actual movement can be selected between 0 and 4 s.

The status of the switches, sensors and drive control modules is displayed on the front panel
of the main control module by a set of LED indicators.
All the switches of the manual control module are duplicated at the front panel of the main control module, 
since the former is located away from the manipulator ($i.e.$ in the control room), and the latter
is mounted in the vicinity of the drives. 

\section{Control software}

\subsection{Hardware control level}
The manipulator communicates with the control PC via an RS232 COM port by sending commands and receiving status messages.
The command contains the key (an ASCII character) and optional operands.
The PC can send to manipulator commands to control vertical ("Y" command), horizontal ("X" command) drives and vacuum pumps ("V" command).
In addition it can query the manipulator status by sending the "Q" command.
The communication speed can be 9.6 or 19.2 kBaud adjustable with the switch at the rear side of the electronic box.
The instruction for the horizontal drive, the vertical drive and the valves consist of 2 bytes. The 1st byte (key) defines the direction or valves, the 2nd byte (operand) defines the position to go to.

\subsection{Algorithmic level : SySal}
\begin{figure}
\includegraphics[width=0.5\textwidth]{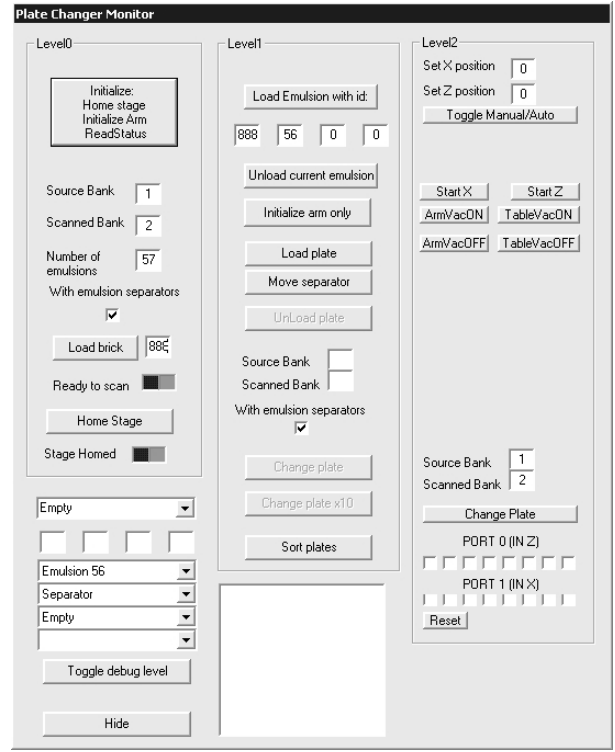}
\caption{The manipulator interactive interface within the SySal emulsion scanning framework.}
\label{fig:sysal}
\end{figure}

The official software framework to perform the routine scan of the OPERA emulsions is called SySal\footnote{https://sysal2000.sa.infn.it}.
It is written in VC++ and works under control of WinXP OS. The framework is written with the use of the COM architecture.
The manipulator interface is implemented as a dynamically linked library (DLL). The DLL provides to the framework following interface methods: LoadPlate(film identifier) and UnloadPlate().
DLL also provides visual interactive interface to perform brick/emulsion load/unload operations as well as 
low level operations, such as arm motions, which may be used for debugging purposes (Figure~\ref{fig:sysal}).

\subsection{Algorithmic level : ROOT}
An alternative framework more suitable for development purposes has been developed as a part of the FEDRA\footnote{http://ntslab01.na.infn.it/fedra/} emulsion reconstruction 
framework used for data analysis in OPERA. This package is written using ROOT\footnote{http://root.cern.ch} conventions
and libraries. The manipulator control is implemented as an object named AcqPlateChanger inherited from standard ROOT ancestor
 TObject.
In addition to the main methods LoadBrick(), UnloadBrick(), LoadPlate() and UnloadPlate(), a set of low level commands is implemented.
The RootCINT C++ interpreter allows the operator to communicate to the object interactively, while the 
scanning supervisor program is realized as a ROOT script.

\section{Performance}

\subsection{Positioning precision}

\begin{figure}
\includegraphics[width=0.5\textwidth]{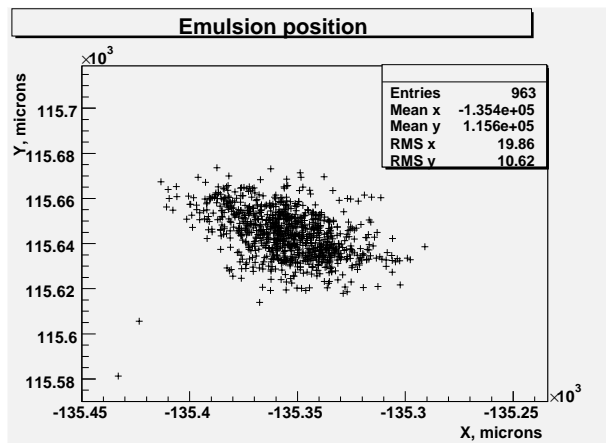}
\caption{Emulsion placement precision: position of the emulsion film fiducial mark in the coordinate system of the table.}
\label{fig:pospres}
\end{figure}

The emulsion film placement accuracy is determined by the following contributions.
The film in the bank box is positioned loosely with a certain gap. This gap 
is defined by the tolerances on the emulsion size, which, according to producer (Fuji Film) specifications, is 100~$\mu$m for each transverse dimension.
The bank size corresponds to the maximum emulsion film size. Therefore, an average emulsion has the freedom up to 100~$\mu$m.
This value limits the absolute positioning accuracy. 
However, relative positioning accuracy or positioning repeatability
can be much better. On the "X" axis it is mainly defined by the arm rigidity and the horizontal drive positioning accuracy.
On the "Y" axis the main contribution comes from the elasticity of the vacuum cups.

Placement precision was measured by repeatedly placing of one emulsion film from the bank to the microscope table,
each time measuring the emulsion film position by finding special mark printed on it by automatic scanning microscope.
The mark position measurement accuracy is better than 1~$\mu$m, so it does not contribute to the result.
The measured mark coordinates are plotted in Figure~\ref{fig:pospres}. 
The accuracy on the "Y" axis is about 10~$\mu$m and on the "X" axis about 20~$\mu$m due to arm vibration.
Maximum (peak-to-peak) relative displacement is well within 100~$\mu$m.
This precision largely satisfies the OPERA requirements to have placement accuracy
better than one microscope view (about $390\times310$~$\mu$m$^2$).

\subsection{Operation speed}
\begin{figure}
\includegraphics[width=0.5\textwidth]{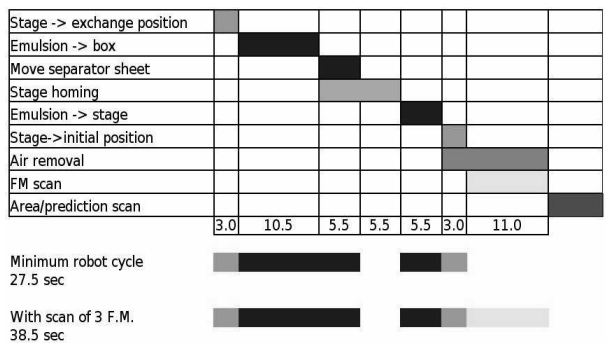}
\caption{The time sequence of the emulsion film placement.
(F.M. - fiducial mark)
}
\label{fig:timing}
\end{figure}

The manipulator operation is composed of a number of elementary actions:
move the arm horizontally to predefined position $X_i$,
move the arm vertically to predefined position $Y_i$,
switch the arm vacuum on, switch the table vacuum off, etc.
The combinations of these basic steps form three main algorithmic actions of a higher level:
\begin{enumerate}
\item{Move film from bank A to bank B}  
\item{Move film from microscope to bank A}
\item{Move film from bank B to microscope}
\end{enumerate}
The total time needed for the complex operation of emulsion film exchange is shown in Figure~\ref{fig:timing}.
The operations shown in the figure are:
\begin{description}
\item[\texttt{Stage->Exchange position}] Moving the microscope object table into the predefined position, where it is accessible to the manipulator arm.
\item[\texttt{Emulsion->Box :}] Moving the emulsion film from the microscope object table to the scanned bank box.
\item[\texttt{Move separator sheet :}] Moving the plastic separator sheet from the source bank box to the scanned bank box.
\item[\texttt{Stage homing :}] Optional operation of finding the reference position for the microscope object table.
\item[\texttt{Emulsion->Stage :}] Moving the emulsion film from the source bank box to the microscope object table.
\item[\texttt{Stage->Initial position :}] Moving the microscope object table back to scanning position.
\item[\texttt{Air removal :}] Time needed to evacuate the air from between the emulsion film and the glass surface of the microscope object table.
\item[\texttt{FM scan :}] Searching and measuring positions of the fiducial marks to establish the reference coordinate system.
\item[\texttt{Area/Prediction scan :}] Main track scan operations.
\end{description}

High placement accuracy reduces the duration of the fiducial marks scan down to 11.0 s so that the full operation cycle
time is about 40 s per emulsion film.

\section{Safety issues}
The automatic plate manipulator belongs to a pick-and-place robot family and is characterized by the limited and well defined space
occupied by all moving parts that are potentially harmful. The force that the manipulator motors apply to the moving parts is limited 
 to provide safety in case the manipulator would meet an obstacle.
This upper limit is chosen to be 1 kg to make it safe for human operators arms accidentally put in the way of the manipulator.

The motion of the manipulator is preceeded by sound and blinking light warning signals, so that the operator has about 4 seconds 
to react and clear the manipulator space.

Electrical safety is characterized by standard requirements to equipment powered by 220V AC with a 
current consumption below 10A and operating at a relative humidity below $70\%$. 

\section{Conclusions}
Present experiments in the field of neutrino physics, such as the OPERA neutrino oscillation experiment,
require high tracking accuracy and large detector mass at the same time. The recent developments in the emulsion
film production allowed to satisfy these requirements. The newly designed automatic emulsion scanning systems 
are able to perform the emulsion scanning at an outstanding speed of 20 $cm^2$ per hour. 
However the film feeding speed has become a bottleneck.

A novel automatic computer controlled emulsion manipulator for modern automatic high-speed scanning stations has been developed for the special purposes of the OPERA experiment.
In this paper we have shown, that the process of feeding emulsion films to the scanning microscope has been successfully automatized,
allowing to process a large number of emulsion films without human intervention.

The performace required by the projected scanning load of OPERA has been successfully achieved: emulsion placement 
accuracy is well within 100~$\mu$m peak-to-peak, the emulsion replacing time is about 30 seconds.
Five manipulators have been installed and currently operate at the European scanning laboratories of the OPERA collaboration
and provide 24 hours a day unmanned scanning process with a very low failure rate.

\section{Acknowledgments}
We acknowledge the members of the OPERA collaboration for suggestions and discussions.
We gratefully acknowledge the technical staff of the Laboratory for High Energy Physics
of the University of Bern, in particular S. Lehman, H. Ruetsch, F. Nydegger and J.-C. Roulin for the invaluable support in
constructing the manipulator. We thank L. Martinez for the IT support during the development, and S. Gamper, who
has made his diploma work on this subject. We also would like to thank A. Ereditato for cooperation in the work on the paper. 
We express our thankfulness to the Swiss National Foundation for the financial support.

\end{document}